\documentclass[sigconf]{acmart}

\usepackage{enumitem}
\usepackage{bm}
\usepackage{multirow} 
\usepackage{etoolbox}
\usepackage{adjustbox}
\usepackage{tcolorbox}
\usepackage{colortbl}
\usepackage{geometry}
\usepackage{balance}
\usepackage{amsmath}
\usepackage{algorithm}
\usepackage{algorithmic}
\usepackage{float}

\usepackage{extarrows}

\DeclareMathOperator*{\argmax}{arg\,max}

\newcommand{\ie}{\emph{i.e., }}

\AtBeginDocument{%
  }

\setcopyright{acmlicensed}
\copyrightyear{2018}
\acmYear{2026}
\acmDOI{XXXXXXX.XXXXXXX}
\acmConference[ACM KDD '26]{32nd SIGKDD Conference on Knowledge Discovery and Data Mining}{August 09-13, 2026}{Jeju, Korea}

\acmISBN{978-1-4503-XXXX-X/2018/06}

\begin{document}
\begin{sloppypar}
\title{Align-for-Fusion: Harmonizing Triple Preferences via Dual-oriented Diffusion for Cross-domain Sequential Recommendation}

\author{Yongfu Zha}
\email{zhayongfu@nudt.edu.cn}
\affiliation{%
  \institution{National University of Defense Technology}
  \city{Changsha}
  \country{China}
}


\author{Xinxin Dong}
\email{dongxinxin@nudt.edu.cn}
\affiliation{%
  \institution{National University of Defense Technology}
  \city{Changsha}
  \country{China}
}

\author{Haokai Ma}
\email{haokai.ma1997@gmail.com}
\authornotemark[1]
\thanks{*Corresponding author}
\affiliation{%
 \institution{National University of Singapore}
 \city{Singapore}
 \country{Singapore}}

\author{Yonghui Yang}
\email{yyh.hfut@gmail.com}
\affiliation{%
  \institution{National University of Singapore}
  \city{Singapore}
  \country{Singapore}}

\author{Xiaodong Wang}
\email{xdwang@nudt.edu.cn}
\affiliation{%
  \institution{National University of Defense Technology}
  \city{Changsha}
  \country{China}}




\renewcommand{\shortauthors}{Trovato et al.}

\begin{abstract}
Personalized sequential recommendation aims to predict the appropriate items to users from their behavioral sequences. To alleviate the data sparsity and interest drift issues, conventional approaches typically utilize the additional behaviors from other domains via cross-domain transition. However, existing cross-domain sequential recommendation (CDSR) algorithms follow the \emph{align-then-fusion} paradigm which conducts the representation-level alignment across multiple domains and mechanically combine them for recommendation, overlooking the fine-grained multi-domain fusion. Inspired by the advancements of diffusion models (DMs) in distribution matching, we propose an \emph{align-for-fusion} framework for CDSR to \textbf{H}arm\textbf{o}nize t\textbf{ri}ple preferences utili\textbf{z}ing Dual-\textbf{o}rie\textbf{n}ted DMs (HorizonRec). Specifically, we first investigate the uncertainty injection of DMs and attribute the fundamental factor of the instability in existing DMs recommenders to the stochastic noise and propose a Mixed-conditioned Distribution Retrieval strategy which leverages the retrieved distribution from users’ authentic behavioral logic as a bridge across the triple domains, enabling consistent multi-domain preference modeling. To suppress the potential noise and emphasize target-relevant interests during multi-domain user representation fusion, we further propose a Dual-oriented Preference Diffusion method to guide the extraction of preferences aligned with users’ authentic interests from each domain under the supervision of the mixed representation. We conduct extensive experiments and analyses on four CDSR datasets from two distinct platforms to verify the effectiveness and robustness of our HorizonRec and its effective mechanism in fine-grained fusion of triple domains. Source codes and corresponding datasets will be released upon acceptance.
\end{abstract}

\ccsdesc[500]{Information systems~Recommender systems}

\keywords{Recommender System, Cross-domain Sequential Recommendation, Diffusion Models}



\maketitle
\section{Introduction}
\label{section1}

\begin{figure}[!t]
    \centering
    \includegraphics[width=0.98\linewidth]{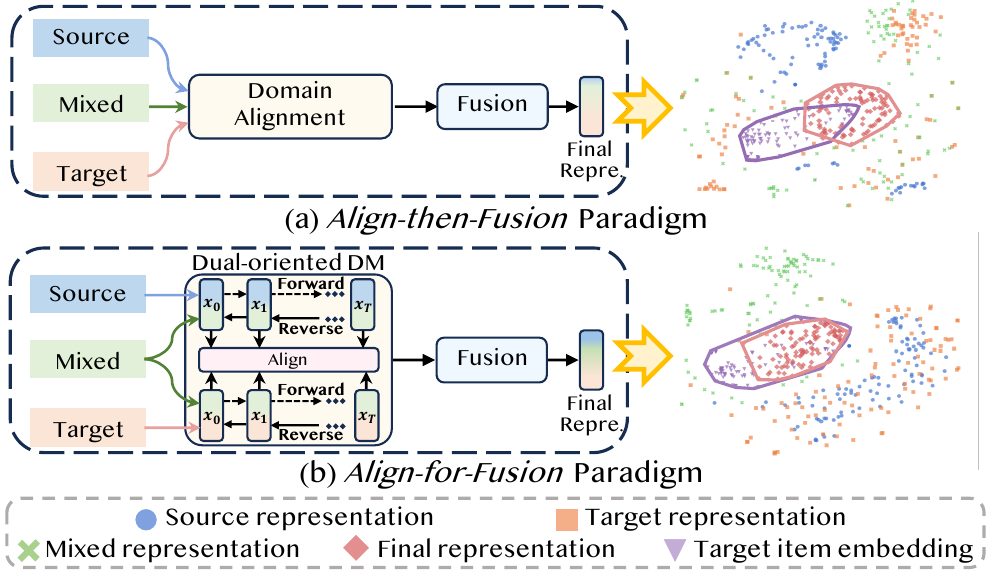}
    \vspace{-0.2cm}
     \caption{Comparison of the illustration and the T-SNE visualization on user representations and target item embeddings on Music$\rightarrow$Book setting between the \emph{Align-then-Fusion} and \emph{Align-for-Fusion} paradigms. Our method yields better alignment and target-aware representation.}
    \vspace{-0.4cm}
    \label{fig.motivation}
\end{figure}

Sequential recommendation (SR) aims to capture users' dynamic interest from their behavioral sequences and provide appropriate items to them \cite{SASRec,CL4SRec}. However, the data sparsity problem and interest drift issue inherent in SR hinder its temporal preference understanding, ultimately limiting its performance \cite{DreamRec,ma2024negative}. Recently, cross-domain sequential recommendation (CDSR) has attracted increasing attention, as it can leverage users' sequential behaviors from auxiliary domains to enrich the target domain and infer the dynamic preferences via interest shifts observed elsewhere \cite{Tri-CDR,C2DSR}.

Drilling down into this topic, prior studies typically focus on the knowledge transfer from source to target domain, particularly in cross-domain share-account recommendation \cite{piNet,DAGCN}. Recently, a series of studies highlight the significance of mixed domain which encompasses user behaviors from both source and target domain and facilitate preference modeling via temporal alignment \cite{ABXI} and multi-granular correlation modeling \cite{Tri-CDR, C2DSR}. As illustrated in Figure~\ref{fig.motivation} (a), these works primarily follow the \emph{align-then-fusion} paradigm to separate the representation-level alignment across multiple domains with the subsequent mechanical combination. To verify this issue, we also conduct the T-SNE visualization on Tri-CDR \cite{Tri-CDR} with 100 randomly-selected and their domain-specific representations, final representations, and target item embeddings in the inference phase. We can observe the noticeable overlap and intersection among representations from triple domains, yet the distribution of final representation exhibits a significant gap from that of target items—an inevitable consequence of the \emph{align-then-fusion} paradigm. It may inadvertently introduce noise interference from other domains or discard critical domain-specific signals, resulting in the suboptimal alignment with users' true interest distributions in target domain.

Inspired by the remarkable performance of Diffusion Models (DMs) in Computer Vision and Natural Language Processing, some researchers have initiated a line of pioneering works to explore their potential in recommendation. For instance, DiffuRec \cite{DiffuRec} and DreamRec \cite{DreamRec} successfully inject the uncertainty modeling capabilities of DMs into sequential recommendation, but their focus remains limited to intra-domain settings, making them less applicable to cross-domain scenarios. More recently, several cross-domain recommendation endeavors (\ie DiffRec \cite{DiffRec} and DMCDR \cite{DMCDR}) utilize DMs to capture the inter-domain collaborative correlations, overlooking the sequential patterns inherent in user behaviors. Therefore, when directly applied to CDSR tasks that require joint modeling of cross-domain and temporal dependencies, they yield suboptimal performance.

To alleviate these issues, we propose an \emph{align-for-fusion} framework, HorizonRec, which \emph{regards mixed domain as a unified Horizon to break through the boundaries of cross-domain preference modeling, enabling Dual-oriented DM to precisely harmonize user preferences across triple behavioral landscapes}. Specifically, we begin by analyzing the noise interference and information absence issues in \emph{align-then-fusion} paradigm, analyze the critical factor of instability in existing DM-based recommenders and attribute it to the deviation introduced by random noise. Motivated by the core of Retrieval-Augmented Generation, we propose a Mixed-Conditioned Distribution Retrieval (MDR) module to retrieve distributions most relevant to users' global interests based on their authentic behavioral logic. These retrieved signals serve as the initial distribution for the subsequent diffusion process to ensure that user preference representations remain globally aligned throughout diffusion, thereby facilitating the proposed \emph{align-for-fusion} paradigm. Then we propose a Dual-oriented Preference Diffusion (DPD) module to simultaneously conduct diffusion on source and target domains under the guidance of the mixed domain, enabling the extraction of domain-specific knowledge that is closely aligned with the user's global preference and target-domain interest. As illustrated in Figure~\ref{fig.motivation} (b), the final representation is entirely located within the distributional region of target item embedding in HorizonRec, demonstrating the effectiveness of our \emph{align-for-fusion} paradigm. It enables dual-oriented diffusion to serve as a horizon for harmonizing the aligned preferences across triple domains, ultimately yielding a unified representation of the users' personalized interests.

We have conducted extensive experiments on four CDSR settings from two diverse platforms to demonstrate the superiority of our HorizonRec. We also conduct various ablation studies, parameter sensitive analysis, computational complexity analysis and visualization to verify the effectiveness of our proposed MDR and DPD. The contributions are summarized as follows:
\begin{itemize}[leftmargin=*, topsep=0.2pt,parsep=0pt]
    \item We have delicately verified the significance of \emph{align-for-fusion} paradigm for target-related knowledge extraction and fine-grained triple preferences harmonizing. To the best of our knowledge, we are among the pioneer to investigate this paradigm via dual-oriented diffusion models in CDSR.
    \item We design a Mixed-conditioned Distribution Retrieval module which retrieves the globally-focused distribution to support diffusion process for stable multi-domain preference modeling.
    \item We conduct extensive experiments and analyses to demonstrate the effectiveness of each component within our HorizonRec on various datasets.
\end{itemize}
\section{Related Work}
\subsection{Sequential Recommendation}
Sequential Recommendation (SR) aims to model the users' dynamic preference pattern with the historical behavior sequences and recommend the appropriate item to him/her~\cite{liao2025mitigating,CPRec}. Early SR approaches attempt to infer users' temporal preference with the Markov Chains \cite{MC1, MC2}, Convolution Neural Network \cite{Caser}, and Recurrent Neural Network \cite{GRU4Rec}. With the success of the self-attention mechanism, SASRec \cite{SASRec} leverages Transformer as the sequential encoder to model users' short- and long-term preferences. CL4SRec \cite{CL4SRec} additionally designs several contrastive learning tasks to achieve mutual information maximization on sequence level. Recently, PDRec \cite{PDRec}, DiffuRec \cite{DiffuRec}, and DreamRec \cite{DreamRec} utilize the advanced Diffusion Models to inject the uncertainty into user preference modeling in a generative manner to improve its robustness and generalization. 
However, the research emphasis of these studies remains confined to uncovering the intra-domain dynamic preference patterns, hindering their effectiveness when directly applied to cross-domain scenarios.

\subsection{Cross-domain Recommendation}
Prior cross-domain recommendation (CDR) studies \cite{Gromov,DTCDR,esrlycdr,wang2024making} aim to enhance target-domain performance by leveraging auxiliary-domain information. For instance, SSCDR \cite{SSCDR} introduces a semi-supervised mapping method to transfer preferences across domains. UniCDR \cite{UniCDR} employs domain masking and contrastive learning to extract domain-invariant representations. Recently, some works start introducing diffusion models into CDR, where DiffCDR\cite{DiffCDR} and DMCDR \cite{DMCDR} leverage it to model preference uncertainty and enhance cross-domain alignment. Extending to cross-domain sequential recommendation (CDSR), Tri-CDR \cite{Tri-CDR} designs a triple contrastive objective to capture fine-grained correlations among triple domains. SyNCRec \cite{SyNCRec} recognize the harmful gradients to mitigate negative transfer, and ABXI \cite{ABXI} extracts domain-invariant components for adaptive integration into domain-specific representations.
However, these works mechanically align inter-domain user behaviors via tailored optimization objectives, while overlooking the nuanced correlations and distinctions among multi-domain features during the fusion phase.

\begin{figure*}[!t]
    \centering
    \includegraphics[width=0.98\linewidth]{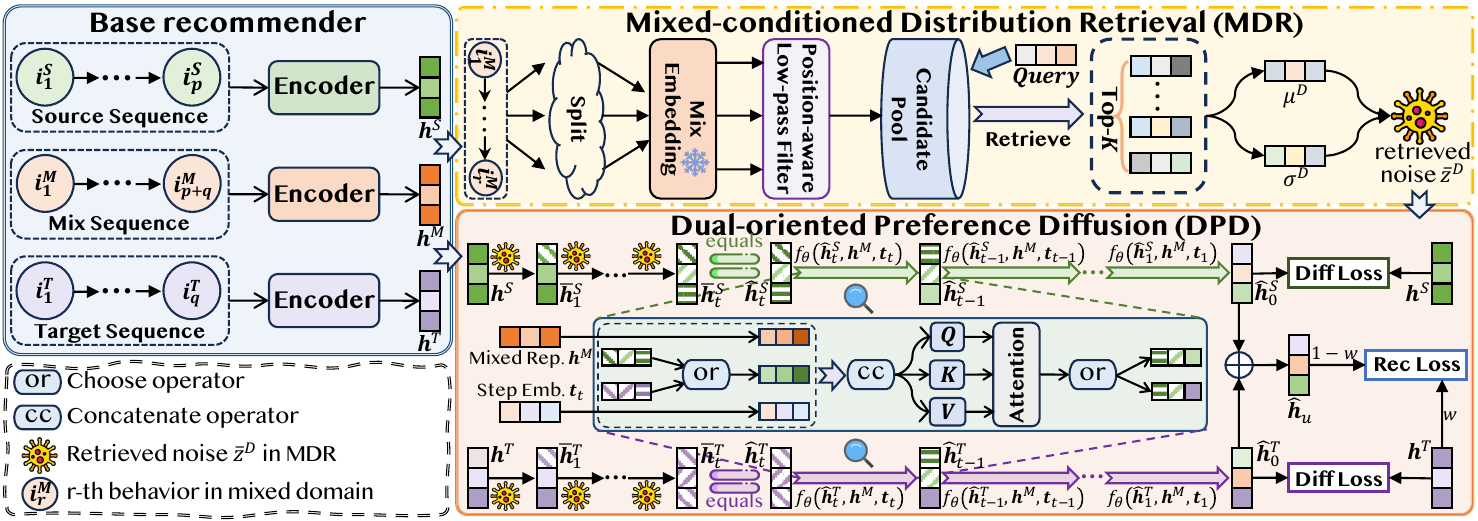}
    \vspace{-0.4cm}
    \caption{The overview of our proposed HorizonRec framework.}
    \label{fig.HorizonRec}
    \vspace{-0.5cm}
\end{figure*}

\subsection{Diffusion Models for Recommendation}
Motivated by the remarkable achievement of Diffusion Models (DMs) in image synthesis \cite{Diff4Image1,Diff4Image2} and categorical data modeling \cite{Diff4cg}, some researchers attempt to incorporate it into recommendation \cite{Recdiff,nips24diff,Seedrec}. For instance, DiffRec \cite{DiffRec} models the generative process of user interactions via denoising. MCDRec \cite{MCDRec} introduces a multi-channel diffusion process to enhance user-item semantic alignment. 
To further investigate its distribution matching property, recent studies attempt to apply it in SR and CDR tasks, which focus more on evolving user preferences. DiffuRec \cite{DiffuRec} generates future behaviors through a diffusion decoder, guided by a preference-oriented reverse process. DreamRec \cite{DreamRec} perturbs temporal sequences with noise and leverages conditional diffusion to generate denoised behavior sequences. For the more specific CDR tasks, the pioneering DiffCDR \cite{DiffCDR} utilize DM for multi-domain preference sharing, incorporating auxiliary-domain behaviors during decoding. DMCDR \cite{DMCDR} further proposes a preference-guided diffusion strategy that conditions the generation on cross-domain preferences, improving representations for cold-start users.
However, existing works focus either on the multi-modality of items or the temporal preference shifts within a single domain, while rarely model the cross-domain and sequential dependencies from users' cross-domain behaviors.

\section{Preliminary}
\label{sec.preliminary}
Inspired by the powerful distribution matching ability of Denoising Diffusion Probabilistic Models (DDPM) \cite{DDPM}, we adopt it as the diffusion backbone for our HorizonRec. Here, we briefly introduce its key components~\footnote{The complete details are provided in Appendix~\ref{prove} due to space constraints.}.

\noindent
\textbf{Forward Process.} DDPM defines a pre-specified Markov process that progressively adds Gaussian noise to a clean input $\mathbf{x}_0$, such that the final state $\mathbf{x}_T$ approximates an isotropic Gaussian distribution: $\mathbf{x}_T \sim \mathcal{N}(\mathbf{0}, \mathbf{I})$.
\begin{equation}
q(\mathbf{x}_t|\mathbf{x}_{t-1}) = \mathcal{N}(\mathbf{x}_t; \sqrt{1 - \beta_t} \, \mathbf{x}_{t-1}, \beta_t \mathbf{I}).
\end{equation}
By unfolding the forward process, the diffusion state distribution at time step $t$ can be formulated as:
\vspace{-0.1cm}
\begin{equation}
q(\mathbf{x}_t|\mathbf{x}_0) \!=\! \mathcal{N}(\mathbf{x}_t; \!\sqrt{\bar{\alpha}_t} \mathbf{x}_0, \!(1 - \bar{\alpha}_t) \mathbf{I}), \ \text{where} \quad \bar{\alpha}_t = \prod_{s=1}^t (1 - \beta_s).
\vspace{-0.1cm}
\end{equation}
Through reparameterization technology, $\mathbf{x}_t$ can be represented as $\mathbf{x}_t = \sqrt{\bar{\alpha}_t} \, \mathbf{x}_0 + \sqrt{1 - \bar{\alpha}_t} \, \boldsymbol{\epsilon}, \quad \boldsymbol{\epsilon} \sim \mathcal{N}(\mathbf{0}, \mathbf{I}).$

\noindent
\textbf{Reverse Process.} Regarding the reverse process, we utilize the original representation $\mathbf{x}_0$ as the reverse target to generate the estimated representation $\hat{\mathbf{x}}_0$ via a estimator $f_\theta(.)$ which is conditioned on the noised input $\mathbf{x}_t$ and the additional side information:
\begin{equation}
\hat{\mathbf{x}}_0 \!=\! f_\theta(\mathbf{x}_t, \mathbf{c}, t)\!=\!f_\theta(\sqrt{\bar{\alpha}_t} \, \mathbf{x}_0 + \sqrt{1 - \bar{\alpha}_t} \, \boldsymbol{\epsilon}, \mathbf{c}, t),
\end{equation}
where $\mathbf{c}$ denotes the specific contextual guidance (e.g., the mixed-domain representation in our HorizonRec).

\noindent
\textbf{Training Objective.} To align with the reverse process, we use the Mean-Squared Error (MSE) loss function on the original representation $\mathbf{x}_0$ and the estimated representation $\hat{\mathbf{x}}_0$ to optimize the estimator $f_\theta(.)$:
\begin{equation}
\mathcal{L} = \mathbb{E}_{\mathbf{x}_0, \boldsymbol{\epsilon}, t} \left[
\left\| \mathbf{x}_0 - f_\theta \left( \sqrt{\bar{\alpha}_t} \, \mathbf{x}_0 + \sqrt{1 - \bar{\alpha}_t} \, \boldsymbol{\epsilon}, \mathbf{c}, t \right) \right\|^2 \right].
\end{equation}
This formulation enables stable training and encourages the generative model to recover semantically meaningful user representations guided by cross-domain context.

\section{Methodology}
\subsection{Problem Formulation} 

CDSR aims to predict users' next interaction in the target domain by exploiting dependencies within and across domains. To achieve this, we first define the \emph{source behavioral sequence} as $S_{u}^{S}\!=\!(i_{1}^{S}, i_{2}^{S}, \ldots, i_{p}^{S})$ and \emph{target behavioral sequence} as $S_{u}^{T}\!=\!(i_{1}^{T}, i_{2}^{T}, \ldots, i_{q}^{T})$ of user $u$, where $p$ and $q$ denote the sequence lengths of source domain $S$ and target domain $T$, $i_{k}^{S}\!\in\!\mathcal{I}^{S}$ and $i_{j}^{T}\!\in\!\mathcal{I}^{T}$ represent the $k$-th and the $j$-th behavior in $S_{u}^{S}$ and $S_{u}^{T}$, respectively. Then we construct the \emph{mixed behavioral sequence} $S_{u}^{M} \!=\!\left(i_{1}^{M}, i_{2}^{M}, \ldots, i_{ p + q}^{M}\right)$ by chronologically merging $S_u^S$ and $S_u^T$. Here, CDSR attempts to maximize the probability of the next behavior $i_{q+1}^{T}$ of user $u$ in target domain according to his/her historical sequences $S_{u}^{S}$, $S_{u}^{M}$, and $S_{u}^{T}$, which can be formulated as:
\vspace{-0.2cm}
\begin{equation}
\argmax_{i_{q+1}^{T} \in \mathcal{I}^{T}} p(i_{q+1}^{T} \mid S_{u}^{S}, S_{u}^{T}, S_{u}^{M}).
\label{cdsr_objective}
\end{equation}
\vspace{-0.4cm}

\subsection{Overall Framework}
In this section, we introduce a \emph{align-for-fusion} framework, HorizonRec, which aims to harmonize triple domains via dual-oriented diffusion to explicitly capture and integrate users' preferences from the source, target, and mixed domains. As illustrated in Figure~\ref{fig.HorizonRec}, our HorizonRec includes two key modules: \romannumeral1) Mixed-conditioned Distribution Retrieval, which aims to dynamically regulate the noise distribution in the diffusion process by incorporating historical behaviors from mixed domain that are close to users' global preferences. By introducing structured external priors, it controls the direction and magnitude of noise injection through similar sequences, suppressing irrelevant domain information and enhancing the accuracy of CDSR. \romannumeral2) Dual-oriented Preference Diffusion: We introduce a dual-oriented diffusion models on source and traget domain simultaneously, harmonizing users' local interests with the global preference strategy and facilitating joint modeling of user preferences across triple domains.

\subsection{Base Recommender}
CDSR aims to predict users' next behavior in the target domain according to their historical behavior sequences from other domains, relying on the accurate modeling of temporal behavior patterns within each domain.
To achieve this, we employ SASRec~\cite{SASRec} as our domain-specific recommender.
Taking the target domain $T$ as an example, we first map the behavioral sequence $S_u^{T} = (i_1^T, i_2^T, \ldots, i_q^T)$ into the embedding sequence $\mathbf{E}_u^{T} = (\mathbf{e}_1^T, \mathbf{e}_2^T, \ldots, \mathbf{e}_q^T)$ using the embedding matrix $\mathbf{E}^T \in \mathbb{R}^{|\mathcal{I}^T| \times d}$. Then we utilize the self-attention mechanism and pointwise feed-forward network to obtain the hidden sequence $\bm{H}_u^T$:
\begin{equation}
\text{Attn}_u^T = \text{Softmax}\left(\bm{Q}^T(\bm{K}^T)^\top/\sqrt{d}\right)\bm{V}^T.
\end{equation}
where the hidden sequence is computed as $\bm{H}_u^T \!=\!\text{ReLU}(\text{Attn}_u^T\bm{W}_1\!+\!\bm{b}_1)\bm{W}_2 \!+\! \bm{b}_2$,  $\bm{Q}^T, \bm{K}^T, \bm{V}^T$ are projected from the embedding sequence $\mathbf{E}_u^{T}$, $\bm{W}_1$, $\bm{W}_2$, $\bm{b}_1$, $\bm{b}_2$ are weights and biases, respectively. The hidden sequences $\bm{H}_u^S$ and $\bm{H}_u^M$ of user $u$ on the source and mixed domains are similarly constructed through this process. 
Following the representative CDSR method \cite{Tri-CDR}, we utilize SASRec (S+T) as our base cross-domain sequential recommender. Specifically, we concatenate the sequence-level representations $\bm{h}_u^S$ and $\bm{h}_u^T$ of source and target domains to generate the final user representation $\bm{h}_u$ via a multi-layer perceptron (MLP)~\footnote{To reduce notational clutter, we omit user $u$ in the following sections for simplicity.}. 

\subsection{Mixed-conditioned Distribution Retrieval}
\label{sec:mdr}
Conventional DM-based recommenders typically inject user-agnostic Gaussian noise into user behaviors~\cite{DiffRec} or item representations~\cite{MCDRec}, where the controlled noise can facilitate the smoother representations modeling, thereby partially alleviating data sparsity. However, such purely random noise may disrupt subtle cross-domain correlations, exacerbate distributional shifts between domains, and obscure the consistency of user preferences across triple domains in CDSR tasks. To address this issue, we propose the Mixed-conditioned Distribution Retrieval (MDR) module, which constructs behavior-aligned and user-specific noise from historical mixed-domain interactions to support the following diffusion process within our HorizonRec.

Given the mixed-domain sequence $S_u^M = (i_1^M, i_2^M, \ldots, i_{|S_u^M|}^M)$ of user $u$, we firstly extract all the contiguous subsequences ending with target-domain items to form the retrieval candidate:
\begin{equation}
    \mathcal{D}_u = \left\{ (i_1^M, \ldots, i_l^M) \mid 1 < l \leq |S_u^M|,\; i_l^M \in \mathcal{I}^T \right\}.
\end{equation}
Then, we extract the representations of each item within the subsequence $d \in \mathcal{D}_u$ from the pre-trained item embedding matrix and then utilize a position-aware low-pass filter~\cite{Butterworth} to highlight the more recent behaviors in candidate embeddings $\mathbf{e}_d$.
\begin{equation}
\label{eq.MDR}
\mathbf{e}_d = \sum_{j=k}^{l} \left( c - 1/(1 + \left( \frac{p(i_j)}{l - j + 1} \right)^n) \right) \mathbf{e}(i_j^M),
\end{equation}
where $p(i_j)$ denotes the position of item $i_j$, $c$ controls the initial value of weight of $\mathbf{e}(i_j^M)$, and $n$ represents the decay rate. According to Equation~\ref{eq.MDR}, we can extract all the filtered embeddings from $\mathcal{D}_u$ of user $u$ to construct the user-specific embedding matrix $E_u$. Applying this procedure across all users results in the construction of a global retrieval database as $\mathcal{D} = \bigcup_{u \in \mathcal{U}} E_u^d$.

For each sequence, we utilize user's final representation $\bm{h}^D$ ($D \in \{S, T\}$) from source and target domain to retrieve the top-$K$ most similar segments from $\mathcal{D}$, denoted $\{\mathbf{d}^D_1, \mathbf{d}^D_2, \ldots, \mathbf{d}^D_K \}$. These segments can serve as the semantic anchors to construct the enhanced noise distribution as follows:
\begin{equation}
\bar{z}^D = \mu^D + \sigma^D \boldsymbol{\xi}, \quad \boldsymbol{\xi} \sim \mathcal{N}(\mathbf{0}, \mathbf{I}),
\end{equation}
where $\mu^D \!=\! \frac{1}{K}\sum_{i=1}^K (\mathbf{d}_i \!-\!\bm{h}^D) $ and $\sigma^D$ denote the mean and standard deviation across all the retrieved segments. And the retrieved noise $\bar{z}^D$ is sampled based on the retrieved segments from this user historical behaviors under the specific domain $D$. 

\emph{Theoretical Properties.}
MDR provides behaviorally aligned noise with two desirable properties:
\begin{lemma}[Retrieval Distribution Structure]
\label{lem:retrieval_structure}
The retrieval distribution from MDR satisfies:
\begin{equation}
    \bar{z}^D = \mu^D + \sigma^D \boldsymbol{\xi}, \quad \text{with} \quad {\sigma^D}^2 \leq \sigma_{\text{std}}^2,
\end{equation}
In expectation, and $\mu^D$ points toward target-ending trajectories.
\end{lemma}

\begin{proposition}[Dual-Oriented Property]
\label{thm:mdr_property}
Under the smoothness of the specific CDSR task:
\begin{equation}
    \mathbb{E}[p(i_{q+1}^T|\bm{h}^D + \bar{z}^D, \bm{h}^M)] \geq \mathbb{E}[p(i_{q+1}^T|\bm{h}^D + \bar{z}_{std}, \bm{h}^M)].
\end{equation}
where $\bar{z}_{std} \sim \mathcal{N}(\mathbf{0}, \mathbf{I})$ denotes the standard Gaussian noise.
\end{proposition}
These analyses support the intuition behind the design of MDR, showing that the retrieved noise $\bar{z}^D$ not only provides dual-oriented guidance, but also exhibits lower variance than standard Gaussian perturbations in CDSR tasks.

\subsection{Dual-oriented Preference Diffusion}
\label{sec:dpd}

CDSR assumes that user behaviors from multiple domains reflect diverse and dynamic interests, providing valuable auxiliary signals for target-domain modeling. However, indiscriminately leveraging all behaviors may introduce redundancy or noise, undermining recommendation performance. To mitigate this, we propose a Dual-oriented Preference Diffusion (DPD) module on the basic of the enhanced noise distribution in MDR, which leverages the generative power of diffusion models to uncover shared user intents. It injects retrieval-enhanced noise into source and target domain representations and utilizes mixed-domain preferences as a semantic bridge to guide cross-domain interest modeling and harmonize user preferences among triple domains.

\emph{Dual-oriented diffusion.}
Zooming into the implementation details of existing DM-based cross-domain recommenders, we observe that these approaches typically condition the diffusion process of target-domain representations on source-domain information, while neglecting explicit modeling of user behavioral representations in the source domain. Notably, source-domain behavioral sequences also encode informative dynamic interest patterns that are crucial for preference transfer—particularly for constructing accurate representations in CDSR. Given a user’s domain-specific representation $\bm{h}^D$ ($D \in \{S, T\}$), we inject the noise sampled from the retrieved distribution in MDR as follows:
\begin{equation}
\bar{\bm{h}}^D_t = \sqrt{\bar{\alpha}_t} \bm{h}^D + \sqrt{1 - \bar{\alpha}_t} \, \overline{z}^D.
\end{equation}
where $\overline{z}^D$ denotes the noise retrieved from historical behaviors in domain $D$. This formulation allows both domains to evolve under the guidance of semantical stochasticity.

\emph{Mixed-domain guided reverse.}
To harmonize users' interest patterns within the source and target domains, we introduce the mixed-domain representation $\bm{h}^M$ as a semantic condition. It encodes global user preferences by temporally reorganizing behaviors from both domains, thereby emphasizing domain-specific signals during the forward process in diffusion modeling. During the reverse process, this mixed-domain signal serves as an anchor to conditionally guide the co-evolution of $\bar{\bm{h}}^S_t$ and $\bar{\bm{h}}^T_t$ toward user-intent-aligned directions. Here, the noised representations $\bar{\bm{h}}^D_t$ equals the initial reconstructed representation $\hat{\bm{h}}^D_t$ according to the standard DDPM framework. Therefore, we concatenate the reconstructed representations with the mixed-domain embedding $\bm{h}^M$ and the step embedding $\bm{t}_e$ of step $t$,to generate the reconstructed representation $\hat{\bm{h}}_{t-1}^D$ of step $t-1$:
\begin{equation}
\hat{\bm{h}}_{t-1}^D=f_\theta(\bar{\bm{h}}^D_t, \bm{h}^M, \bm{t}_t) = \mathrm{Softmax}\left( \bm{Q} \bm{K}^\top/\sqrt{d} \right) \bm{V},
\end{equation}
where $\bm{Q}$, $\bm{K}$, and $\bm{V}$ are query, key, and value matrices obtained via learnable linear projections from $\text{Concat}(\bar{\bm{h}}^D_t, \bm{h}^M, \bm{t}_e)$. The iterative reverse process can be formulated as follows:
\begin{equation}
\hat{\bm{h}}_{t}^D\!\xlongrightarrow{f_\theta\!(\hat{\bm{h}}^D_t, \bm{h}^M, \bm{t}_t)\!}\!\hat{\bm{h}}_{t\!-\!1}^D\!\xlongrightarrow{f_\theta\!(\bar{\bm{h}}^D_{t\!-\!1}, \bm{h}^M, \bm{t}_{t-1})\!}\!\cdots\!\xlongrightarrow{f_\theta\!(\hat{\bm{h}}^D_1, \bm{h}^M, \bm{t}_1)\!}\hat{\bm{h}}_{0}^D.
\end{equation}

\begin{proposition}[Mixed-domain Alignment Advantage]
\label{thm:mixed_guidance}
The conditional mechanism of estimator $f_\theta(\cdot)$ should satisfy:
\begin{equation}
    d_{\text{align}}(\hat{\bm{h}}_0^S, \hat{\bm{h}}_0^T | \bm{h}^M) \leq d_{\text{align}}(\bm{h}^S, \bm{h}^T).
\end{equation}
where $d_{\text{align}}(.)$ denotes the cross-domain correlation.
\end{proposition}

\emph{Diffusion loss.}
Following the DDPM objective in Section~\ref{sec.preliminary}, we optimize $f_\theta$ by minimizing the reconstruction error, enabling the reverse process to recover domain-specific preferences consistent with the semantics of the mixed domains.
\begin{equation}
\mathcal{L}_{\text{diff}} \!=\! \mathbb{E}_{\bm{h}^D} \!\left\| \bm{h}^D - \hat{\bm{h}}_{0}^D \right\|^2, \ \text{where} \ \hat{\bm{h}}_{0}^D\!=\!f^{(0 \leftarrow t)}_\theta(\bar{\bm{h}}^D_t, \bm{h}^M, \bm{t}_t).
\label{eq:ldiff}
\end{equation}

This DPD module leverages interest-related noise retrieved by MDR within the dual-oriented DM to uncover domain-specific user preferences, while MDR in turn benefits from the co-evolution of source and target representations within DPD.
\begin{proposition}[Convergence Guarantee]
\label{prop:convergence}
Under the Lipschitz continuity of $f_\theta(.)$, the dual-oriented DM is expected to converge into a distribution $p_\infty$ that satisfies the following condition:
\begin{equation}
    \mathcal{W}_2(p_\infty, p_{\text{target}}) \leq \mathcal{W}_2(p_\infty^{\text{std}}, p_{\text{target}}).
\end{equation}
where $\mathcal{W}_2$ is the 2-Wasserstein distance and $p_\infty^{\text{std}}$ denotes the convergence under standard Gaussian noise.
\end{proposition}

\subsection{Training Strategy}

After completing the reverse process in DPD, we can obtain the constructed user representations of source and target domains as $\hat{\bm{h}}_0^S$ and $\hat{\bm{h}}_0^T$, and then we fuse them to generate the intermediate user representation as $\hat{\bm{h}}_u = \hat{\bm{h}}_0^S + \hat{\bm{h}}_0^T$.
To preserve the target-specific semantics essential for recommendation, we integrate it with the original target-domain representation $\bm{h}_u$ to generate the final user representation $\tilde{\bm{h}}_u$ as follows:
\begin{equation}
    \tilde{\bm{h}}_u = (1 - w)\hat{\bm{h}}_u + w\bm{h}_u,
\end{equation}
where $w \in [0,1]$ controls the contribution of target-domain information in the fused user representation.

Then we calculate the score by matching the final user representation $\tilde{\bm{h}}_u$ and the candidate item embeddings. Here, our HorizonRec is optimized with cross-entropy loss:
\begin{equation}
    \mathcal{L}_{\text{rec}} = \frac{1}{|\mathcal{U}|} \sum_{u \in \mathcal{U}} -\log \frac{\exp(\tilde{\bm{h}}_u^\top \bm{e}_{n+1})}{\sum_{j \in \mathcal{I}^T} \exp(\tilde{\bm{h}}_u^\top \bm{e}_j)},
\end{equation}
where $\bm{e}_{n+1}$ denotes the embedding of the ground-truth item.

Finally, the overall training objective combines the recommendation loss and the diffusion loss:
\begin{equation}
    \mathcal{L} = \mathcal{L}_{\text{rec}} + \lambda \mathcal{L}_{\text{diff}}.
\end{equation}
where $\lambda$ denotes the weight of $\mathcal{L}_{\text{diff}}$.

\subsection{Complexity Analysis}
We further analyze the additional computational costs introduced by HorizonRec to explore its scalability in real-world industry. 
Specifically, it first retrieve the top-$K$ similar sequences from the database $\mathcal{D}$ through similarity computation in the MDR module, incurring the time complexity of $\mathcal{O}(|\mathcal{D}| \times d)$. Regarding the DPD module, it conducts the conditional attention across triple domains with the time complexity as $\mathcal{O}(T \times d^2)$, where $T$ and $d$ denote the diffusion steps and embedding dimensions, respectively. 
To sum up, our HorizonRec introduces only an additional time cost of $\mathcal{O}(T \times d^2 + |\mathcal{D}| \times d)$ during training. Compared to traditional CDSR methods such as Tri-CDR, which incurs a time complexity of $\mathcal{O}(3L^2 \times d)$ due to its tri-domain self-attention mechanism (where $L$ denotes the maximum sequence length), HorizonRec achieves an asymptotically similar time complexity.
Talking about the space complexity , the primary costs arise from storing the retrieval database $\mathcal{D}$, which requires $\mathcal{O}(|\mathcal{D}| \times d)$, and maintaining the intermediate diffusion states during training, which adds an additional $\mathcal{O}(T \times d)$. Notably, HorizonRec benefits from the pre-computed retrieval database during the inference phase to avoid repeated database construction, enabling its efficiency in online serving. Meanwhile, it is able to maintain the desirable performance even with a small number of steps $T$, making it practically feasible for real-world CDSR applications.
\section{Experiments}
\label{sec:experiments}
We conduct extensive experiments and analyses to answer the following research questions:\\
\textbf{RQ1}: How does HorizonRec perform compared to state-of-the-art sequential, cross-domain, and cross-domain sequential recommendation baselines? (see Section~\ref{performance_comp})\\
\noindent
\textbf{RQ2}: What are the contributions of different components in HorizonRec? (see Section~\ref{ablation})\\
\noindent
\textbf{RQ3}: How effectively does HorizonRec align and integrate user preferences from triple domains? (see Section~\ref{indepth_anlysis})\\
\noindent
\textbf{RQ4}: How robust is HorizonRec with respect to key hyperparameters, and what is its computational efficiency compared to existing competitive baselines? (see Section~\ref{robustness_analysis})\\

\begin{table}[!t]
\caption{Detailed statistics of our CDSR settings.}
\vspace{-0.2cm}
\centering
\label{dataset}
\begin{tabular}{c|cc|cc}
\toprule
Dataset & \multicolumn{2}{c}{Amazon Toy \& Game} & \multicolumn{2}{|c}{Douban Book \& Music} \\
\midrule
Domain & Toy & Game & Book & Music \\
Users & 7,994 & 7,994 & 6,367 & 6,367 \\
Items & 37,840 & 11,735 & 50,448 & 36,504 \\
Interactions & 113,605 & 82,855 & 189,971 & 181,570  \\
Density &0.0376\% &0.1211\% &0.0591\% &0.0781\% \\
\bottomrule
\end{tabular}
\vspace{-0.3cm}
\end{table}

\begin{table*}[!t]
\caption{Performance comparison of HorizonRec and baselines on Douban Book$\rightarrow$Music and Douban Music$\rightarrow$Book. All relative improvements are statistically significant ($p < 0.01$) under paired t-tests over five random seeds.}
\vspace{-0.2cm}
\label{tab:douban_cross_domain}
\centering
\resizebox{\linewidth}{!}{ 
\begin{tabular}{c|cccccc|cccccccc}
\toprule 
\multirow{2}{*}{Algorithms} & \multicolumn{6}{c|}{Douban Book$\rightarrow$Music} & \multicolumn{6}{c}{Douban Music$\rightarrow$Book} \\
& HR@5 & NDCG@5 & HR@10 & NDCG@10 & HR@20 & NDCG@20 & HR@5 & NDCG@5 & HR@10 & NDCG@10 & HR@20 & NDCG@20 \\
\midrule
GRU4Rec &0.0439	&0.0228	&0.0753	&0.0313	&0.1109	&0.0401	&0.0498	&0.0246	&0.0786	&0.0349	&0.1112	&0.0436\\
SASRec &0.0468	&0.0241	&0.0773	&0.0335	&0.1145	&0.0431	&0.0532	&0.0281	&0.0821	&0.0373	&0.1151	&0.0451 \\
CL4Rec &0.0523	&0.0281	&0.0821	&0.0389	&0.1156	&0.0479	&0.0586	&0.0356	&0.0913	&0.0468	&0.1211	&0.0549 \\
Diffurec &0.0545	&0.0333	&0.0786	&0.0411	&0.1123	&0.0496	&0.0612	&0.0384	&0.0942	&0.0491	&0.1279	&0.0577 \\
\midrule
SSCDR &0.0523	&0.0312	&0.0845	&0.0389	&0.1136	&0.0489	&0.0632	&0.0398	&0.0965	&0.0479	&0.1336	&0.0589 \\
UniCDR &0.0609	&0.0397	&0.0955	&0.0501	&0.1301	&0.0588	&0.0739	&0.0456	&0.1078	&0.0541	&0.1418	&0.0646 \\
DMCDR &0.0606	&0.0392	&0.0948	&0.0495	&0.1296	&0.0583	&0.0741	&0.0463	&\underline{0.1101}	&\underline{0.0562}	&\underline{0.1465}	&\underline{0.0658} \\
\midrule
C2DSR &0.0578	&0.0356	&0.0898	&0.0456	&0.1201	&0.0542	&0.0718	&0.0446	&0.1025	&0.0503	&0.1389	&0.0633 \\
Tri-CDR &0.0605	&0.0387	&0.0955	&0.0506	&0.1341	&0.0601	&0.0742	&0.0463	&0.1083	&0.0538	&0.1443	&0.0655 \\
SyNCRec &0.0628	&0.0392	&\underline{0.0986}	&\underline{0.0513}	&\underline{0.1386}	&\underline{0.0608}	&\underline{0.0746}	&0.0461	&0.1089	&0.0549	&0.1436	&0.0649 \\
ABXI &\underline{0.0633}	&\underline{0.0398}	&0.0978	&0.0506	&0.1355	&0.0596	&0.0739	&\underline{0.0468}	&0.1093	&0.0551	&0.1445	&0.0652 \\ \midrule
HorizonRec &\textbf{0.0710} &\textbf{0.0440} &\textbf{0.1067} &\textbf{0.0555} &\textbf{0.1469} &\textbf{0.0655} &\textbf{0.0797} &\textbf{0.0491} &\textbf{0.1166} &\textbf{0.0604} &\textbf{0.1543} &\textbf{0.0698}  \\
Rel.Impro. &\color{red}12.16\% &\color{red}10.55\% &\color{red}8.22\% &\color{red}8.19\% &\color{red}5.99\% &\color{red}7.73\% &\color{red}6.84\% &\color{red}4.91\% &\color{red}5.90\% &\color{red}7.47\% &\color{red}5.32\% &\color{red}6.08\% \\ \bottomrule
\end{tabular}}
\vspace{-0.3cm}
\end{table*}

\subsection{Experimental Settings}
\subsubsection{Dataset}
\label{appendix:preprocess}
We conduct experiments on four real-world cross-domain recommendation scenarios from two platforms: Amazon Toy$\rightarrow$Game, Amazon Game$\rightarrow$Toy, Douban Book$\rightarrow$Music and Douban Music$\rightarrow$Book. 
We randomly sample users who have interaction histories in both the source and target domains. Following the filtering strategy of Tri-CDR~\cite{Tri-CDR}, we retain users with at least three interactions in each domain and chronologically sort their behaviors within each domain based on the timestamps.
To model the cross-domain behavioral dynamics, we construct \emph{mixed-domain sequences} by merging source and target interactions in temporal order. For evaluation, we follow the protocol of DiffuRec~\cite{DiffuRec}: the penultimate interaction in the target domain is used for validation, and the last interaction is held out for testing. Dataset statistics are summarized in Table~\ref{dataset}.

\subsubsection{Baselines}
\label{appendix:baselines}
We compare HorizonRec against four single-domain sequential recommenders, three cross-domain recommenders, and four cross-domain sequential recommenders to comprehensively demonstrate its effectiveness.

\noindent
\textbf{1) Single-domain Sequential Recommendation.}
We utilize GRU4Rec~\cite{GRU4Rec}, SASRec~\cite{SASRec}, CL4SRec~\cite{CL4SRec}, and DiffuRec~\cite{DiffuRec}. These methods model user preferences from temporal patterns within a single domain using techniques such as recurrent networks, self-attention, contrastive learning, and diffusion-based modeling.

\noindent
\textbf{2) Cross-domain Recommendation.}
We evaluate SSCDR~\cite{SSCDR}, UniCDR~\cite{UniCDR}, and DMCDR~\cite{DMCDR}, which transfer knowledge across domains by learning domain-invariant or aligned representations. While SSCDR and UniCDR use shared latent spaces and self-supervised objectives, DMCDR introduces diffusion at the embedding level without modeling sequential dynamics.

\noindent
\textbf{3) Cross-domain Sequential Recommendation.}
We compare with C2DSR~\cite{C2DSR}, Tri-CDR~\cite{Tri-CDR}, SyNCRec~\cite{SyNCRec}, and ABXI~\cite{ABXI}. These methods jointly model sequential behaviors across domains by leveraging contrastive alignment, multi-task learning, synchronization strategies, or adversarial training to enhance cross-domain preference extraction.

\subsubsection{Implementation Details}
\label{ed}
All experiments are conducted on an NVIDIA RTX 4090 with Python 3.10.9. 
For fair comparison, we initialize both HorizonRec and baselines that support pretrained embeddings with the same domain-specific pretrained embeddings from the source, target, and mixed domains, while other methods are trained from scratch following their original implementations~\footnote{Implementation details of all baselines are provided in Appendix~\ref{appendix:baselines}.}. 
We adopt SASRec~\cite{SASRec} as the base sequence encoder, the batch size, embedding dimension and max length are set as 512, 64 and 200 for all algorithms, respectively. 
The weight $w$ of original target-domain representation and the weight $\lambda$ for controlling diffusion loss are both selected from $\{0.1, 0.2, \ldots, 0.9\}$, and the number of diffusion steps is set as 32.
In the retrieval-augmented diffusion process, the sliding window length $l$ is capped at 200. The early-position importance factor $c$ is set to 1.5, and the distance decay rate $n$ is 2. We vary the number of retrieved similar sequences $K$ from $\{5, 10, 15\}$. To reduce the randomness, we conduct five times for each experiment with different random seeds and report the average results.

\begin{table*}[!t]
\caption{Performance comparison of HorizonRec and baselines on Amazon Game$\rightarrow$Toy and Amazon Toy$\rightarrow$Game. All relative improvements are statistically significant ($p < 0.01$) under paired t-tests over five random seeds.}
\vspace{-0.2cm}
\label{tab:amazon_cross_domain}
\centering
\resizebox{\linewidth}{!}{ 
\begin{tabular}{c|cccccc|cccccccc}
\toprule
\multirow{2}{*}{Algorithms} & \multicolumn{6}{c|}{Amazon Game$\rightarrow$Toy} & \multicolumn{6}{c}{Amazon Toy$\rightarrow$Game} \\
& HR@5 & NDCG@5 & HR@10 & NDCG@10 & HR@20 & NDCG@20 & HR@5 & NDCG@5 & HR@10 & NDCG@10 & HR@20 & NDCG@20 \\ \midrule
GRU4Rec &0.0276	&0.0178	&0.0368	&0.0216	&0.0483	&0.0245 &0.0496	&0.0306	&0.0712	&0.0379	&0.1056	&0.0461 \\
SASRec &0.0339	&0.0228	&0.0413	&0.0253	&0.0514	&0.0271 &0.0541	&0.0349	&0.0758	&0.0417	&0.1079	&0.0497 \\
CL4Rec &0.0362	&0.0241	&0.0445	&0.0288	&0.0526	&0.0298 &0.0564	&0.0362	&0.0782	&0.0436	&0.1113	&0.0526 \\
Diffurec &0.0335 &0.0301	&0.0375	&0.0324	&0.0411	&0.0342 &0.0495 &0.0385	&0.0631	&0.0414	&0.0878	&0.0492\\ \midrule
SSCDR &0.0344	&0.0277	&0.0436	&0.0279	&0.0529	&0.0288 &0.0587	&0.0388	&0.0816	&0.0438	&0.1135	&0.0548\\
UniCDR &\underline{0.0391}	&0.0309	&\underline{0.0492}	&\underline{0.0339}	&\underline{0.0575}	&\underline{0.0361}  &0.0655 &0.0483	&0.0908	&0.0561	&0.1235	&0.0638	\\
DMCDR &0.0346	&0.0281	&0.0446	&0.0313	&0.0537	&0.0309 &0.0627	&0.0461	&0.0841	&0.0527	&0.1131	&0.0588	\\
\midrule
C2DSR &0.0381	&0.0298	&0.0474	&0.0328	&0.0545	&0.0339 &0.0649	&0.0471	&0.0892	&0.0549	&0.1203	&0.0619	\\
Tri-CDR &0.0389	&\underline{0.0311}	&0.0481	&0.0337	&0.0553	&0.0345 &0.0646	&0.0478	&0.0896	&0.0556	&0.1215	&0.0623	\\
SyNCRec  &0.0368	&0.0296	&0.0467	&0.0318	&0.0542	&0.0329 &\underline{0.0668}	&\underline{0.0491}	&\underline{0.0921}	&\underline{0.0578}	&\underline{0.1265}	&\underline{0.0651}	\\
ABXI &0.0366	&0.0298	&0.0471	&0.0321	&0.0545	&0.0333 &0.0656	&0.0483	&0.0916	&0.0566	&0.1234	&0.0641	\\ \midrule
HorizonRec &\textbf{0.0421} &\textbf{0.0330} &\textbf{0.0517} &\textbf{0.0361} &\textbf{0.0610} &\textbf{0.0383}  &\textbf{0.0688} &\textbf{0.0522} &\textbf{0.0949} &\textbf{0.0602} &\textbf{0.1299} &\textbf{0.0680} \\
Rel.Impro. &\color{red}7.67\% &\color{red}6.11\% &\color{red}5.08\% &\color{red}6.49\% &\color{red}6.09\% &\color{red}6.09\% &\color{red}2.99\% &\color{red}6.31\% &\color{red}3.04\% &\color{red}4.15\% &\color{red}2.69\% &\color{red}4.45\% \\ \bottomrule
\end{tabular}}
\vspace{-0.1cm}
\end{table*}

\subsection{Performance Comparison (RQ1)}
\label{performance_comp}
We conduct experiments on four CDSR settings to evaluate the effectiveness of our HarizonRec under standard top-$k$ metrics: HR@$k$ and NDCG@$k$ at $k=5$, $10$, and $20$. Here, we highlight the best result in \textbf{bold}, and the second-best in \underline{underlined} in Table~\ref{tab:douban_cross_domain} and Table~\ref{tab:amazon_cross_domain} and observe the following discoveries:

(1) HorizonRec consistently outperforms all baselines across domains and evaluation metrics, validating the effectiveness of incorporating denoising diffusion into cross-domain sequential recommendation. In particular, the retrieval-enhanced noise injection (MDR) and the dual-oriented denoising process (DPD) jointly enable HorizonRec to model fine-grained cross-domain dependencies and recover latent user intents more accurately than conventional methods.

(2) Compared with single-domain sequential recommenders (e.g., GRU4Rec, SASRec, CL4Rec, DiffuRec), HorizonRec achieves consistent and substantial gains. These baselines model only intra-domain temporal dependencies and neglect auxiliary domains, which limits their ability to generalize user intent. While contrastive learning in CL4Rec and diffusion modeling in DiffuRec offer marginal improvements, their lack of cross-domain alignment and interaction modeling restricts their effectiveness in CDSR settings.

\begin{figure*}[!t]
    \centering
    \includegraphics[width=0.98\linewidth]{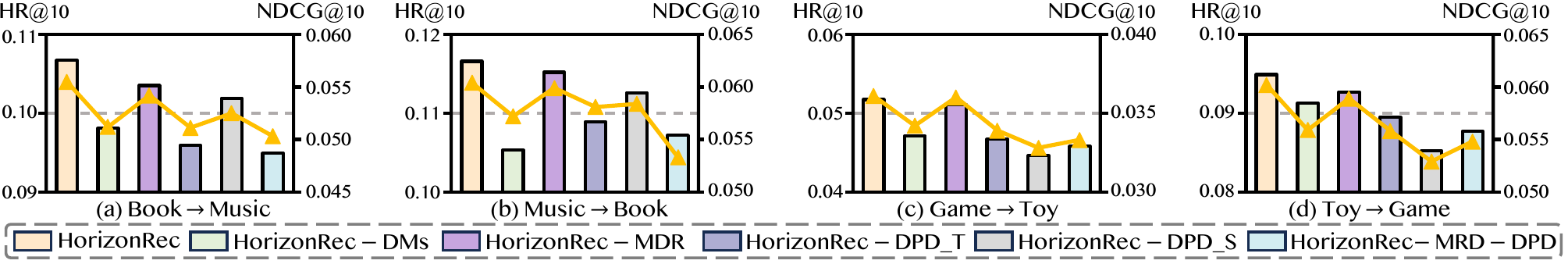}
    \vspace{-0.2cm}
    \caption{Ablation results on four CDSR settings. It verifies the effectiveness of \emph{align-for-fusion} pipeline and all modules.}
    \vspace{-0.3cm}
    \label{fig.ablation}
\end{figure*}

(3) Compared with cross-domain recommenders that lack explicit sequence modeling (e.g., SSCDR, UniCDR, DMCDR), HorizonRec delivers superior performance. While these models focus on learning a shared representation space across domains, they neglect temporal dependencies—an essential aspect of user behavior in CDSR tasks. In particular, although DMCDR incorporates diffusion-based alignment, it omits sequential structure, which weakens its ability to model evolving user preferences.

(4) Compared with advanced CDSR models (e.g., Tri-CDR, SyNCRec, ABXI), HorizonRec shows robust improvements. While these methods incorporate domain alignment and temporal modeling, their static fusion strategies or adversarial objectives struggle under domain imbalance and noisy conditions. In contrast, HorizonRec’s iterative denoising framework produces more informative and adaptable representations, particularly in sparse settings such as Book$\rightarrow$Music and Toy$\rightarrow$Game.

\subsection{Ablation Study (RQ2)}
\label{ablation}
To investigate the effectiveness of each component within HorizonRec, we conduct a comprehensive ablation study. As shown in Figure~\ref{fig.ablation}, the results confirm that each module plays a crucial and complementary role in enhancing cross-domain recommendation performance.

(1) We compare three variants to evaluate the impact of diffusion across domains. \emph{HorizonRec-DPD\_S} applies denoising only to the source domain, while \emph{HorizonRec-DPD\_T} applies it only to the target domain. Both suffer noticeable performance drops, indicating that one-sided diffusion limits the model’s ability to capture comprehensive preferences. \emph{HorizonRec-DPD}, which disables source-side diffusion entirely, performs the worst, highlighting the importance of symmetric dual-domain denoising for effective alignment.

(2) \emph{HorizonRec-MDR} replaces retrieval-based noise with standard Gaussian noise. While retaining some benefits over baselines, the absence of behaviorally grounded perturbations leads to degraded performance, confirming that MDR’s structured priors provide better semantic guidance than uninformative noise.

(3) \emph{HorizonRec-DMs} removes the diffusion mechanism, relying instead on static representations. The resulting drop shows that iterative denoising is essential for extracting robust user preferences from noisy or incomplete inputs.

(4) \emph{HorizonRec-DPD-MDR} disables both DPD and MDR, using only simple concatenation for fusion. It performs the worst, underscoring the limitations of naive aggregation and the superiority of attention-based fusion, even without denoising.

(5) Overall, each module contributes to HorizonRec’s performance. DPD enables dual-domain alignment, MDR injects semantically informed noise, and the diffusion process supports progressive refinement. Their synergy within the \emph{align-for-fusion} framework establishes a robust foundation for cross-domain representation learning.

\begin{figure}[!t]
    \centering
    \includegraphics[width=1\linewidth]{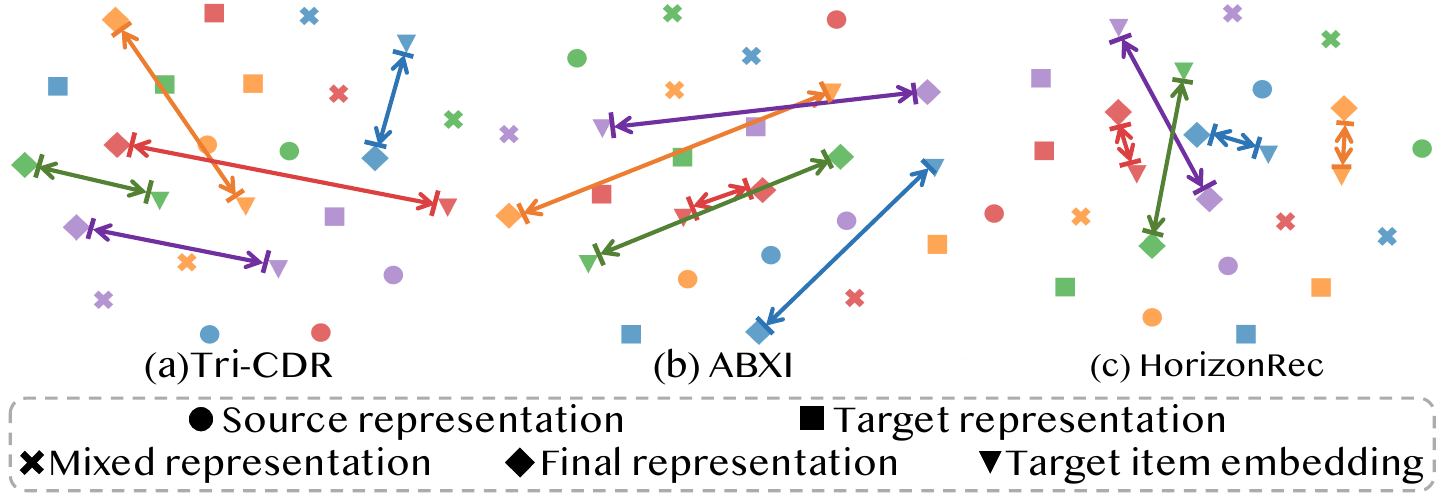}
    \vspace{-0.2cm}
    \caption{T-SNE visualization of two representative CDSR algorithms and our HorizonRec on the Music$\rightarrow$Book setting. We utilize colors to differentiate different users while using shapes to distinguish the multiple representations and target item embedding of the same user.}
    \vspace{-0.3cm}
    \label{fig.5user}
\end{figure}

\subsection{Estimation of Triple Preference Correlation (RQ3)}
\label{indepth_anlysis}
\begin{figure*}[!t]
    \centering
    \includegraphics[width=0.98\linewidth]{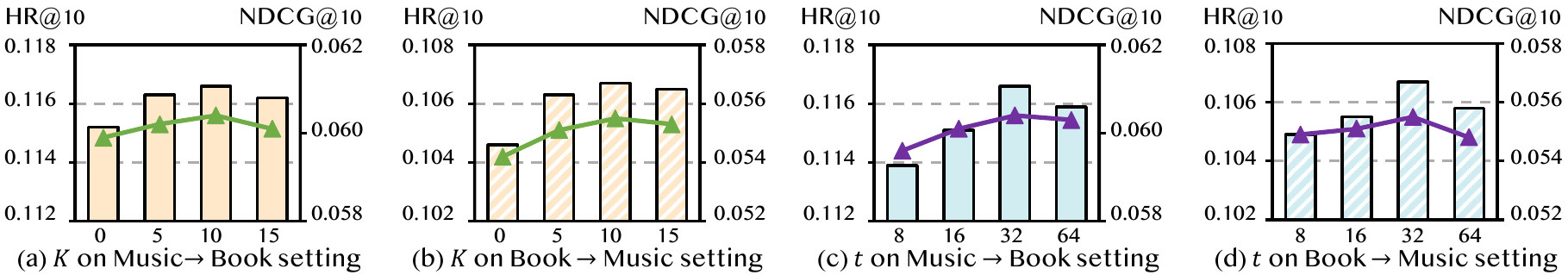}
    \vspace{-0.2cm}
    \caption{Results of parameter sensitivity analysis to investigate the impact of the number of retrieved sequences $K$ and diffusion steps $t$ of our HorizonRec on the Douban Music$\rightarrow$Book \& Book$\rightarrow$Music settings.}
    \vspace{-0.3cm}
    \label{fig.douban_k_t}
\end{figure*}

\begin{figure}[!t]
    \centering
    \includegraphics[width=1\linewidth]{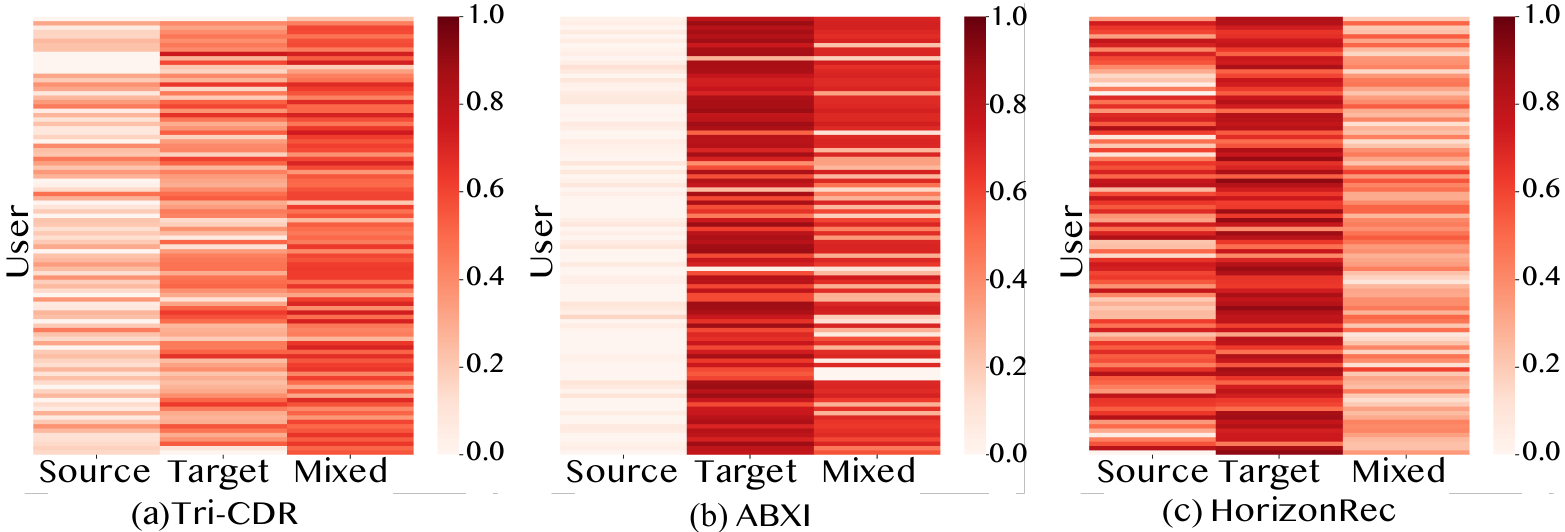}
    \vspace{-0.4cm}
    \caption{Heatmap visualization of the cosine similarity between final and domain-specific representations of two representative CDSR algorithms and our HorizonRec on the Music$\rightarrow$Book setting.}
    \vspace{-0.3cm}
    \label{fig.tsne}
\end{figure}

\subsubsection{Visualization of Cross-Domain Matching}
To further assess the effectiveness of cross-domain preference alignment, we visualize the t-SNE projections of five randomly selected users on the Music$\rightarrow$Book setting, as shown in Figure~\ref{fig.5user}. Different colors represent individual users, and distinct shapes correspond to source, target, mixed-domain, final user, and target item embeddings. Arrows point from each final user representation to its corresponding target item, where shorter arrows suggest stronger alignment and better recommendation quality. As seen in Figure~\ref{fig.5user}, ABXI (b) produces long arrows and disorganized representation layouts, indicating suboptimal alignment and weak cross-domain modeling due to its sequential \emph{align-then-fusion} design. Tri-CDR (a) yields slightly shorter arrows but still exhibits overlapping user clusters, reflecting entangled representations and limited user specificity. In contrast, HorizonRec (c) shows the shortest arrows and most well-separated clusters, revealing that its final user representations are both highly aligned with target semantics and strongly user-discriminative. These results highlight the advantage of our \emph{align-for-fusion} strategy, where MDR provides meaningful, behavior-driven perturbations and DPD enables effective dual-domain refinement through symmetric diffusion.

\subsubsection{Visualization of Cross-Domain Alignment}
To further evaluate the alignment capabilities of different models, we visualize the cosine similarities between each user's final representation and their corresponding source, target, and mixed-domain embeddings, as shown in Figure~\ref{fig.tsne}. Subfigures (a)--(c) present heatmaps for 100 randomly selected users under the Music$\rightarrow$Book setting using Tri-CDR, ABXI, and HorizonRec, respectively.
Tri-CDR (a) produces relatively uniform similarity distributions across all domains, suggesting weak domain-specific discrimination and insufficient modeling of cross-domain interactions. ABXI (b) shows strong alignment with the target domain but poor similarity to the source, indicating limited utilization of auxiliary domain signals. In contrast, HorizonRec (c) achieves consistently high similarities with both source and target domains, along with more distinct user-wise patterns. Its strong alignment with mixed-domain representations further confirms the benefit of the proposed \emph{align-for-fusion} strategy in unifying preferences while preserving domain semantics.

\subsection{Robustness Analysis (RQ4)}
\label{robustness_analysis}
\subsubsection{Parameter Sensitivity Analysis}
We analyze the impact of retrieval number $K$ and diffusion step $t$ on the Douban Music \& Book dataset (Figure~\ref{fig.douban_k_t}). For $K$, increasing the number of retrieved sequences initially improves performance via richer semantic guidance, but overly large $K$ introduces noise and slight degradation. This is because marginally relevant or conflicting sequences begin to dominate, diluting the target-driven perturbation signal. When $K = 0$, the model degenerates to standard Gaussian diffusion and performs worst, indicating that a moderate $K$ balances informativeness and robustness. For $t$, performance peaks at $t=32$, as finer denoising helps reconstruction. Larger $t$ yields diminishing returns or over-smoothing, especially when redundant noise is amplified in later steps. This suggests that excessive diffusion may weaken domain-specific signals, emphasizing the need to calibrate $t$ based on data complexity. These results underscore the importance of hyperparameter tuning for stable and expressive cross-domain representations.

\begin{table}[!t]
\caption{Results of the computational complexity analysis on one training epoch, total training phase, and total inference phase on four CDSR settings.}
\vspace{-0.2cm}
\label{tab:runtime}
\centering
\begin{tabular}{c|c|c|c|c}
\toprule
Dataset & Algorithms & Epoch (s) & Total (s) & Inference (s) \\
\midrule
\multirow{3}{*}{\begin{tabular}[c]{@{}c@{}}Game\\ $\downarrow$\\ Toy\end{tabular}} & Tri-CDR & 23.48 & 563.34 & 30.12 \\
                     & ABXI    & 80.02 & 7688.21 & 2.16  \\
                     & HorizonRec & 14.18 & 146.00 & 0.98 \\
\midrule
\multirow{3}{*}{\begin{tabular}[c]{@{}c@{}}Toy\\ $\downarrow$\\ Game\end{tabular}} & Tri-CDR & 15.71 & 760.52 & 27.55 \\
                      & ABXI    & 80.02 & 7688.21 & 2.16 \\
                      & HorizonRec & 9.23  & 89.16  & 0.81 \\
\midrule
\multirow{3}{*}{\begin{tabular}[c]{@{}c@{}}Book\\ $\downarrow$\\ Music\end{tabular}} & Tri-CDR & 45.84 & 1466.88 & 36.86 \\
                       & ABXI    & 235.93 & 26110.74 & 2.65 \\
                       & HorizonRec & 26.25 & 261.60 & 0.73 \\
\midrule
\multirow{3}{*}{\begin{tabular}[c]{@{}c@{}}Music\\ $\downarrow$\\ Book\end{tabular}} & Tri-CDR & 47.20 & 1982.42 & 38.96 \\
                      & ABXI    & 235.93 & 26110.74 & 2.65 \\
                      & HorizonRec & 27.86 & 277.78 & 0.78 \\
\bottomrule
\end{tabular}
\vspace{-0.3cm}
\end{table}

\subsubsection{Computational Complexity Analysis}
Table~\ref{tab:runtime} presents the runtime comparison between HorizonRec and two representative CDSR baselines, ABXI and Tri-CDR, in terms of per-epoch training time, total training time, and inference latency. Across all datasets, HorizonRec consistently demonstrates superior efficiency, achieving significantly lower training and inference costs. 
For instance, on the \textit{Music} dataset, HorizonRec completes training in 261.60 seconds, compared to 1466.88 seconds for Tri-CDR and 26110.74 seconds for ABXI. Similar trends are observed across other domains, where HorizonRec achieves $3\sim10\times$ faster training and $2\sim30\times$ faster inference. 
The consistently low inference latency (under 1 second across all datasets) further highlights HorizonRec's practicality for real-world deployment. These efficiency gains are primarily attributed to HorizonRec’s lightweight dual-domain diffusion design and precomputed retrieval mechanism, which together ensure scalability for large-scale cross-domain sequential recommendation. 
Note that ABXI shares parameters across domains and trains both domains jointly, which results in identical total training times across different settings.
\section{Conclusion}
We propose HorizonRec, a novel \emph{align-for-fusion} framework for cross-domain sequential recommendation. It first employs the MDR module to retrieve user-specific, behaviorally grounded noise from mixed-domain interactions, constructing a semantically consistent perturbation space. Then, the DPD module conducts symmetric diffusion across domains, guided by mixed-domain representations that preserve mutual information and enhance alignment. This two-stage design enables HorizonRec to capture shared user intents and remain robust under sparse or noisy conditions. Extensive experiments on multiple CDSR benchmarks demonstrate its superiority over existing methods, and our theoretical analysis validates the effectiveness of jointly leveraging MDR and DPD for improved alignment and convergence. HorizonRec’s ability to harmonize diverse domain signals through diffusion-based refinement offers a generalizable solution for complex recommendation environments. In future work, we aim to extend HorizonRec to more domains, incorporate pre-trained user/item representations, and explore cold-start and zero-shot scenarios in practical deployments.

\balance
\bibliographystyle{unsrt}
\bibliography{software}
\end{sloppypar}
\end{document}